\documentclass[12pt,tightenlines,eqsecnum,floats,showpacs,nofootinbib,amsmath,amssymb,aps,prd,superscriptaddress]{revtex4-1}

\usepackage{graphicx}
\usepackage{amsmath,amssymb}
\usepackage{hyperref}
\usepackage{color}

\begin{document}
	
\title{Uniqueness of the Fock quantization of scalar fields in a Bianchi I cosmology with unitary dynamics}

\author{Jer\'onimo Cortez}
\email{jacq@ciencias.unam.mx}
\affiliation{Departamento de F\'isica, Facultad de Ciencias, Universidad Nacional Aut\'onoma de M\'exico, M\'exico D.F. 04510, Mexico.}
\author{Beatriz Elizaga Navascu\'es}\email{beatriz.elizaga@iem.cfmac.csic.es}
\affiliation{Instituto de Estructura de la Materia,
CSIC, Serrano 121, 28006 Madrid, Spain.}
\author{Mercedes Mart\'in-Benito}\email{m.martin@hef.ru.nl}
\affiliation{Radboud University Nijmegen, Institute for Mathematics, Astrophysics and Particle Physics,
Heyendaalseweg 135, NL-6525 AJ Nijmegen, The Netherlands.}
\author{Guillermo A. Mena Marug\'an}\email{mena@iem.cfmac.csic.es}
\affiliation{Instituto de Estructura de la Materia,
CSIC, Serrano 121, 28006 Madrid, Spain.}
\author{Javier Olmedo}\email{jao44@psu.edu}
\affiliation{Department of Physics and Astronomy, Louisiana State University,\\
Baton Rouge, Louisiana 70803-4001, USA}
\affiliation{Institute for Gravitation and the Cosmos \& Physics Department, The Pennsylvania State University,\\
University Park, Pennsylvania 16802 USA}
\author{Jos\'e M. Velhinho}
\email{jvelhi@ubi.pt}
\affiliation{Universidade da Beira Interior, Rua Marqu\^es d'\'Avila e Bolama, 6201-001, Covilh\~a, Portugal.}

\begin{abstract}
The Fock quantization of free scalar fields is subject to an infinite ambiguity when it comes to choosing a set of annihilation and creation operators, a choice that is equivalent to the determination of a vacuum state. In highly symmetric situations, this ambiguity can be removed by asking vacuum invariance under the symmetries of the system. Similarly, in stationary backgrounds, one can demand time-translation invariance plus positivity of the energy. However, in more general situations, additional criteria are needed. For the case of free (test) fields minimally coupled to a homogeneous and isotropic cosmology, it has been proven that the ambiguity is resolved by introducing the criterion of unitary implementability of the quantum dynamics, as an endomorphism in Fock space. This condition determines a specific separation of the time dependence of the field, so that this splits into a very precise background dependence and a genuine quantum evolution. Furthermore, together with the condition of vacuum invariance under the spatial Killing symmetries, unitarity of the dynamics selects a unique Fock representation for the canonical commutation relations, up to unitary equivalence. In this work, we generalize these results to anisotropic spacetimes with shear, which are therefore not conformally symmetric, by considering the case of a free scalar field in a Bianchi I cosmology.
\end{abstract}

\maketitle
\newpage

\section{Introduction}\label{sec:intro}

The present era of high-precision cosmology, with the recent observations provided by  the Planck Collaboration \cite{planck}, is imposing strong bounds on the plausible cosmological models that can be used to describe the physics of the early Universe. Indeed, several candidate models have already been ruled out, while some particular ones  are being
favored \cite{planck-inf}, since they present a good balance between the number of parameters required to specify them and the matching of their predictions with observations. Also, the observed power spectrum of temperature anisotropies of the cosmic microwave background (CMB) seems to reveal certain anomalies at large angular scales. This combination of high-precision observations and potential anomalies with respect to the traditional models is an incentive for the exploration of more general and extended scenarios. Actually, the predictions are commonly based on the assumptions of homogeneity and isotropy of the background spacetime, except for small inhomogeneities that explain the small anisotropies of the CMB. In addition, predictions are usually based on classical general relativity, inheriting the well-known initial singularity problem if one extends the evolution backwards in time. In this respect, bouncing matter cosmologies \cite{mbounce} or quantum cosmology models based on nonperturbative quantizations of general relativity, like loop quantum cosmology \cite{AAN,hybrid-obs}, allow one to extend the evolution beyond the region where the singularity is located, according to general relativity, and may provide predictions that fit better with observations. 

Interestingly, one of the extensions that has not been investigated in full detail and which may shed light on the early physics of the Universe concerns the breakdown of the isotropy hypothesis. Although anisotropies are diluted during inflation, there exists the possibility that they might have left some imprint in those scales of the perturbations that are crossing the cosmological horizon today. Cosmological perturbations in anisotropic spacetimes with flat spatial sections of the Bianchi I type were analyzed in Refs. \cite{bianchi1,bianchi2}. These works provide a formulation in terms of gauge-invariant perturbations. The evolution of these perturbations turns out to be dictated by a set of linear, second-order, ordinary differential equations without dissipation. The dynamical system is complicated by the fact that, unlike what happens in the traditional isotropic models, now the equations of motion couple the scalar and tensor modes for each possible wave vector. This coupling shows that the classification of the perturbations in scalar and tensor modes (as well as vector modes that are nonphysical) has a limited use in an expanding universe with shear. Nonetheless, the coupling turns out not to affect the dominant terms in an asymptotic expansion for infinitely large wave numbers of the modes (with the wave number defined as the norm of the wave vector), suggesting that the classification may be helpful in the discussion of the ultraviolet regime of high wave numbers.

One could expect that the coupling between scalar and tensor modes (even if subdominant in the ultraviolet) might introduce nontrivial correlations among them. These correlations can be explicitly quantified through the scalar-tensor two-point functions, and/or described by means of entanglement between the quantum states of the scalar and tensor perturbations. Indeed, the search for entanglement of scalar and tensor modes is  currently under way \cite{entan1,entan2}, adopting a more general scenario where the entanglement is assumed to be generated by an unspecified mechanism. These investigations analyze the consequences of such a quantum entanglement in the spectrum of anisotropies, yielding nontrivial contributions that might be put to the test and compared with observations in future missions.

A correct analysis of the quantum issues and phenomena that affect the perturbations and their dynamics requires special care, since the system possesses infinitely many degrees of freedom. For this kind of system, the uniqueness Stone--von Neumann theorem \cite{svn} cannot be applied and, in consequence, there is no guarantee of a unique representation of the field analog of Weyl's algebra. Fortunately, it is possible to find privileged Fock representations under the requirement of certain physical criteria. For instance, for linear fields propagating in curved spacetimes, the existence of a timelike Killing vector (together with the positivity of the energy \cite{kayb,kayc}) and/or of a sufficient number of symmetries allows one to select a concrete representation compatible with them. This is the case in some special situations, like free field theories in Minkowski or de Sitter spacetimes \cite{wald}. However, in cosmological scenarios, in general, there are neither timelike Killing vectors nor the sufficient number of symmetries. Nevertheless, in the last decade, this issue has been successfully addressed by demanding a physically well-motivated criterion that serves to select a unique family of Fock representations that are unitarily equivalent. This criterion consists of requiring a  unitarily implementable quantum dynamics, as well as invariance under the symmetries of the equations of motion, symmetries that in many situations can be identified with (or simply restricted to) the Killing symmetries of the background. This criterion was first proposed in the context of Gowdy cosmologies \cite{ccm1,ccmv,ccm2,cmv1,ccmv2,cmvS2,cmsv}, which are midisuperspaces that, after reduction, can be regarded as a scalar field propagating in an 1+1 effective spacetime. The analysis of the applicability of the criterion was later extended to the case of test scalar fields propagating in homogeneous and isotropic universes \cite{cmv2,cmov1,cmov2,cmov3,cmov4,fmov,flat,cfmm,cmmv,flat2,sign,cmv3}, a case that is relevant for the quantization of scalar cosmological perturbations. More recently, the criterion was generalized to the study of fermion fields in conformally ultrastatic backgrounds \cite{fermi1,fermi2,fermi3,fermi4}. Notice that the implementation of the dynamics as a unitary operator on the Fock space determined by the vacuum, as required by this criterion, makes the quantum treatment of the field theory compatible with the standard probabilistic interpretation of quantum mechanics, inasmuch as it provides a 
well-defined Schr\"odinger picture.
 
In the case of scalar fields in homogeneous and isotropic cosmologies, the criterion of a unitary quantum dynamics leads to the introduction of a new pair of canonical variables related with the original field and momentum by a time-dependent (or equivalently background-dependent) transformation: the field is rescaled by a time-dependent function, while the conjugate momentum adopts the inverse scaling and includes an additional term that is linear in the field \cite{cmov4}. This new pair (auxiliary in the construction of a representation for the original one \cite{cmv3}) is completely specified by the requirement of unitarity. Indeed, the time-dependent factors that relate the original field variables and the new ones extract a contribution of the dynamics that cannot be implemented unitarily, whereas the remaining nontrivial evolution is implementable as a unitary operator on Fock space. For instance, in cosmological perturbation theory in isotropic scenarios, while the time variation of the comoving curvature perturbation does not admit a unitary implementation, the Mukhanov-Sasaki variable presents a unitary dynamics in conformal time, the two variables being related by means of a rescaling with a time-dependent function. Thus, the separation of the time dependence in a background-dependent (and hence classically varying) part and a (nontrivial) quantum dynamical part turns out to be a key ingredient in the construction of the Fock description of the system if the quantum evolution is to be unitary \cite{cmv3}. 

Although the above-mentioned separation of the time dependence is achieved by means of a field rescaling--which is a local transformation--in more general situations the extraction of the explicitly time-dependent part of the evolution needs a nonlocal canonical transformation. For scalar fields, the nonlocality comes typically from a dependence on the spectrum of the Laplace-Beltrami operator defined on the spatial sections, so that the transformation appears as mode dependent if one introduces an expansion of the field in the eigenmodes of that operator \cite{fmov,cfmm}. For fermions, the explicitly time-dependent part that must be extracted is different for terms associated with different helicities and chiralities, so that the total transformation is not a rescaling of the field as a whole \cite{fermi2,fermi3}.

The considered uniqueness criterion provides a robust arena for the study of physical phenomena of quantum field theories in cosmological spacetimes, particularly for cases where the quantum dynamics and the interaction with the background geometry play a prominent role, as happens with the evolution of quantum perturbations and anisotropies. As a first step towards the analysis of the Fock quantization of perturbations in Bianchi I spacetimes, in this manuscript we will deal with the quantization of a test scalar field minimally coupled to a background of this kind. We will show that, by imposing vacuum invariance under the spatial symmetries and requiring unitary implementability of the quantum dynamics, we can select a unique Fock representation for the canonical commutation relations, up to unitary equivalence. Moreover, as expected, these conditions determine also a unique splitting of the time variation of the scalar field into an explicit time dependence and a genuine and nontrivial quantum evolution. Our analysis then extends to the anisotropic case the study of scalar fields in a homogeneous and isotropic universe with flat spatial sections \cite{flat,flat2}, proving that the proposed uniqueness criterion works perfectly well in spacetimes without conformal symmetry.

Our discussion will follow closely the strategy that was recently introduced for fermion fields in Refs. \cite{fermi1,fermi2,fermi3,fermi4}. This strategy generalizes the studies carried out so far in the literature for scalar fields in isotropic scenarios inasmuch as it investigates simultaneously the freedom in the choice of Fock representation and the freedom in extracting from the field evolution a nontrivial part that is implemented quantum mechanically. This is achieved by considering families of Fock representations with elements that are related dynamically, so that the Bogoliubov transformations between the corresponding sets of annihilationlike and creationlike variables associated with representations in different families are now allowed to be time dependent. Indeed, this dependence can absorb part of the time variation of the field variables and redefine the dynamical evolution that has to be represented as a unitary endomorphism in Fock space. 

More specifically, we will first determine the necessary and sufficient conditions that a Fock representation must satisfy in order to have a vacuum invariant under translations in the spatial directions (the Killing symmetries of Bianchi I spacetimes with compact sections). In this class of invariant Fock representations, we will consider time-dependent families, obtained by combining the dynamical evolution of the canonical field variables with a time-dependent choice of their linear combinations that define the basic annihilation and creation operators. To simplify the analysis of the dynamics, we will introduce a convenient time-dependent canonical transformation of the original field and momentum variables before passing to the Fock description. It will not be local. However, in the isotropic limit, it will coincide with the canonical transformation that is usually adopted for field theories in isotropic cosmologies \cite{fmov}. Then, we will characterize the families of representations in which all elements are related by transformations that admit a unitary implementation. This characterization will provide the admissible Fock representations and the viable choices of nontrivial dynamics in the quantum theory. Finally, we will prove that all these representations are unitarily equivalent, and that the quantum dynamics is fixed up to terms that are not only unitary, but subdominant in the ultraviolet limit. 

The rest of the paper is organized as follows. In Sec. \ref{sec:class} we give a description of the classical setting that we study in this manuscript, including an analysis of the classical field dynamics. The Fock quantization is introduced at the beginning of Sec. \ref{sec:fock}, and then we characterize the conditions that Fock representations must satisfy in order to be compatible with the criterion of symmetry invariance and unitary evolution. Choosing as reference representation that of a massless scalar field case (the simplest one), we prove in Sec. \ref{sec:uni} that all the previously characterized Fock representations are unitarily equivalent. Besides, we show that the condition of unitary implementability indeed determines the quantum evolution, up to ultraviolet subdominant terms. We conclude in Sec. \ref{sec:concl} with a summary of our results. We also include two appendixes with further details on our derivations.

\section{ Classical test scalar field in a Bianchi I background}\label{sec:class}

Let us consider a Bianchi I spacetime with topology $M = \mathbb{I} \times \mathbb{T}^{3}$, where $\mathbb{T}^{3}$ is the three-torus, with coordinates $x^{i}$ ($i=1,2,3$) belonging to $S^1$, and $\mathbb{I}$ is a connected interval of the real line. The metric of this spacetime in its diagonal form can be written as
\begin{align} \label{eq:metric1}
ds^2=g_{\mu \nu} dx^{\mu} dx^{\nu} = -N^2(t)d t^2 + h_{ij}(t)dx^i dx^j, \quad h_{ij}(t)=a_i^2(t)\delta_{ij},
\end{align}
where $a_{i}$ are the directional scale factors and $N$ is the lapse function. In the expression of the spatial metric $h_{ij}$, the repetition of indices does not mean summation. 
A particular case of these cosmological spacetimes is the (compact) flat Friedmann-Lema\^{\i}tre-Robertson-Walker (FLRW) universe, obtained when the three scale factors are equal. For later convenience we introduce the average scale factor that characterizes the volume expansion,
\begin{align}\label{def_a}
a(t) \equiv \left[a_1(t)a_2(t)a_3(t)\right]^{1/3}\,.
\end{align}

We now introduce a real test scalar field $\phi$ with mass $m$ minimally coupled to this geometry. The Lagrangian density of the field is given by
\begin{align}\label{eq:lagr}
\mathcal{L}_{\phi} = \frac{1}{2} \sqrt{-\det(g)}(-g^{\mu \nu} \nabla_{\mu} \phi \nabla_{\nu} \phi - m^{2} \phi^{2}),
\end{align}
where $\nabla_\mu$ means covariant derivative and $\det(g)$ is the determinant of the spacetime metric.
Given a foliation of the spacetime in spacelike hypersurfaces, let $\Sigma$ denote a particular spatial section with unit, future-directed, normal vector $n^\mu$. Then, the derivative of the field in the normal direction is $n^\mu \nabla_{\mu}\phi$,  the canonical momentum $\pi_\phi$ is defined by
\begin{align}
\pi_\phi = \frac{\delta \mathcal{L}_{\phi}}{\delta(\partial_{t}\phi)}=a^{3}n^\mu \nabla_{\mu}\phi,
\end{align}
and the Hamiltonian density is 
\begin{align}
{\cal H}_\phi=\pi_\phi \partial_{t}\phi -{\cal L}_\phi.
\end{align}
In view of the foliation chosen in Eq. \eqref{eq:metric1}, adapted to homogeneity, we have 
\begin{align}
\pi_\phi =  \frac{a^3}{N}\partial_t \phi.
\end{align}
The Hamiltonian density is then
\begin{align}\label{eq:ham}
{\cal H}_\phi(N) = \frac{1}{2}Na^3 \left(\frac{\pi_\phi^2}{a^6}+h^{ij}\partial_i\phi\partial_j\phi+m^2\phi^{2}\right),
\end{align}
where we have used that, for the Bianchi I spacetime, $\nabla_i=\partial_i$, since the spatial sections are flat.

\subsection{Mode expansion}

In order to handle the spatial dependence of the problem, let us expand the field in a basis of eigenmodes of the Laplace-Beltrami operator $\Delta$ defined on the spatial sections. This operator captures the spatial derivatives that appear in the Klein-Gordon equation derived from the Hamiltonian \eqref{eq:ham}. In the current setup of cosmologies with flat spatial sections, we have
\begin{align}
\Delta(t) = h^{ij}(t)\partial_i\partial_j.
\end{align}

Let ${S}=\{\Phi\}$ be the complexification of the vector space of  solutions to the Klein-Gordon equation. This space is naturally equipped with the product (which is not positive definite on $S$)
\begin{align}
 \langle \Phi_1,\Phi_2\rangle=i \int_{\Sigma} \d V \left(\Phi_1^*n^\mu  \nabla_{\mu}\Phi_2-\Phi_2n^\mu  \nabla_{\mu}\Phi_1^*\right),
\end{align}
where $\d V=\sqrt{\det(h)}d^3\vec{x}$ is the volume element on a given spatial section $\Sigma$, and the symbol $*$ denotes complex conjugation. This product is preserved under the evolution of the solutions to the Klein-Gordon equation, namely, it is independent of the choice of spatial section $\Sigma$. In our case, the product simplifies to
\begin{align}\label{inner}
\langle \Phi_1,\Phi_2\rangle=i \frac{a^3(t)}{N(t)} \int_{\mathbb{T}^{3}} d^3 \vec{x}  \left(\Phi_1^* \partial_t\Phi_2-\Phi_2 \partial_t\Phi_1^*\right).
\end{align} On the other hand, since the Cauchy problem is well posed for the (complexified) Klein-Gordon field in the considered Bianchi I cosmologies, the space of solutions $S$ can be equivalently regarded in our model as the complexified space of initial data for the Klein-Gordon equation.

The operator $\Delta(t)$ is essentially self-adjoint on the space of square integrable functions with respect to the measure $\d V$, and it has a discrete spectrum, owing to the compactness of the spatial sections. Its eigenspaces provide irreducible representations of the group consisting of translations in $\mathbb{T}^3$ (interpretable as compositions of rigid rotations in the three principal spatial directions of the tori), which are Killing symmetries of the considered homogeneous spacetimes. The corresponding eigenfunctions are the set of plane waves $e^{i\vec{k}\cdot\vec{x}}$ in  $\mathbb{T}^{3}$, with $\vec{k}=(k_1,k_2,k_3)\in \mathbb{Z}^3$, and its eigenvalues are time dependent.  On each eigenspace the corresponding eigenvalue can be written as 
\begin{align}
\Delta_k(t) = -k^2\sum_{i=1}^3\left[\frac{\tilde k_i}{a_i(t)}\right]^2,  
\end{align}
with $k=\sqrt{k^{2}_{1}+k^{2}_{2}+k^{2}_{3}}$ and $\tilde k_i = {k_i}/{k}$. We can think of $k$ as the norm of the wave vector $\vec{k}$, and $\tilde k_i$ as the components of the unit vector $\vec{\tilde k}$ in the direction of this wave vector. Note that the spectrum is degenerate: there exist different wave vectors with the same eigenvalue. In particular, the eigenmodes $e^{i\vec{k}\cdot\vec{x}}$ and $e^{-i\vec{k}\cdot\vec{x}}$, corresponding to wave vectors of opposite sign, both have the same eigenvalue. 

Since the physical scalar field is real, we consider a basis of real Fourier modes like the one employed in Ref. \cite{flat}. Then, we adopt the decomposition
\begin{align}
\label{rfd1}
\phi(t,\vec{x})&=\frac{1}{(4\pi^3)^{1/2}}\sum_{\vec{k} \in \mathcal{L}_+}\left[q^{(1)}_{\vec{k}}(t)\cos(\vec{k}\cdot\vec{x})
+q^{(2)}_{\vec{k}}(t)\sin(\vec{k}\cdot\vec{x})\right],\\
\pi_\phi(t,\vec{x})&=\frac{1}{(4\pi^3)^{1/2}}\sum_{\vec{k} \in \mathcal{L}_+}\left[p^{(1)}_{\vec{k}}(t)\cos(\vec{k}\cdot\vec{x})
+p^{(2)}_{\vec{k}}(t)\sin(\vec{k}\cdot\vec{x})\right],
\end{align}
where we have ignored the zero mode, labeled by $\vec{k}=(0,0,0)$, and we have defined the lattice
\begin{align}\nonumber
\mathcal{L}_+ &= \{\vec k;k_1>0\}\cup\{\vec k;k_1=0,k_2>0\}\cup\{\vec k;k_1=0=k_2,k_3>0\}.
\end{align}
Restriction to this lattice avoids duplication of the real modes in the Fourier expansion. On the other hand, the exclusion of the (single) zero mode has no influence in our discussion about unitary implementability of the dynamics in the presence of an infinite number  of degrees of freedom. In any case, if we wanted to quantize this mode, we could deal with it separately. Finally, our normalization is such that the introduced real modes provide pairs of canonically conjugate variables,
\begin{align}\label{eq:poisson-realmod}
\{q^{(l)}_{\vec{k}},p^{(\tilde l)}_{\vec{k'}}\}=\delta^{l,\tilde l}\delta_{{\vec{k}},{\vec{k'}}},
\end{align}
where $l,\tilde l=1,2$. 

One can see that the expression of the Hamiltonian in terms of these modes is formed by two identical contributions without interaction between them:
\begin{eqnarray}
{H}_\phi(N) &=&\sum_{l=1}^2  {H}_\phi^{(l)}(N),\\ \label{eq:ham-realmode}
{H}_\phi^{(l)}(N)&=&\frac{N}{2a^3} \sum_{\vec{k} \in \mathcal{L}_+}  \left\{\left(p^{(l)}_{\vec{k}} \right)^2+ a^6\left[\sum_{i = 1}^{3} \left(\frac{k_{i}}{a_{i}}\right)^{2} + m^{2}\right] \left(q^{(l)}_{\vec{k}} \right)^2\right\}.
\end{eqnarray}
In order to alleviate the notation, in what follows we will use $(q_{\vec{k}}, p_{\vec{k}})$  to generically denote any of the pairs $(q^{(1)}_{\vec{k}}, p^{(1)}_{\vec{k}})$ or $(q^{(2)}_{\vec{k}}, p^{(2)}_{\vec{k}})$ whenever it is convenient.

\subsection{Classical canonical transformation}
\label{trans}

The case of the scalar field in homogeneous and isotropic cosmologies proved that the discussion of the unitary implementability of the dynamics can be simplified with a judicious choice of field variables. This choice involves a time-dependent canonical transformation that extracts background functions from the Klein-Gordon field variables. Typically, the new canonical variables for the modes behave dynamically like harmonic oscillators without dissipative terms. Inspired by this isotropic case, we are also going to consider a time-dependent transformation of the field canonical pairs in the current Bianchi I background. However, owing to the lack of isotropy and in contrast to what happens in the mentioned FLRW case, we are going to introduce a transformation that, instead of being just a time-dependent scaling of the configuration field variable, now is also mode dependent. Therefore, it is going to be nonlocal in the cosmological spacetime. 

Let us introduce then the following canonical change, that is mode dependent as announced, but nevertheless respects the spatial Killing symmetries (which are symmetries of the Laplace-Beltrami operator) inasmuch as it does not mix different modes:
\begin{align}\label{canonical}
\begin{pmatrix}
\tilde q_{\vec{k}}  \\ \tilde p_{\vec{k}}
\end{pmatrix}=\mathcal{P}_{\vec{k}}
\begin{pmatrix}
 q_{\vec{k}}  \\  p_{\vec{k}}
\end{pmatrix},\qquad \mathcal{P}_{\vec{k}}=\begin{pmatrix}
\sqrt{b_{\tilde k}} & 0 \\\frac{1}{2}\frac{\dot{b}_{\tilde k}}{b_{\tilde k}\sqrt{b_{\tilde k}}} & \frac{1}{\sqrt{b_{\tilde k}}}
\end{pmatrix}.
\end{align}
Here $b_{\tilde k}$ is defined as
\begin{align}\label{bk}
b_{\tilde k} =a^{3}\sqrt{\sum_{i=1}^3 \left(\frac{\tilde k_i}{a_i}\right)^{2}},
\end{align}
and the dot stands for the derivative with respect to the harmonic time $\tau$, defined as $N(t)dt=a^3(\tau)d\tau$. The function $b_{\tilde k}$ depends on the three scale factors $(a_1,a_2,a_3)$ and on the unit wave vector $\vec{\tilde k}=\vec{k}/k$, but not on the wave number $k$. Besides, in the limit of isotropy we have $b_{\tilde k}=a^2$, and we recover the usual scaling that is employed in the isotropic scenarios \cite{flat}.

The canonical transformation induces the following change in the Lagrangian of the system (before summing over $l=1,2$)
\begin{align}
\sum_{\vec{k} \in \mathcal{L}_+}\dot q^{(l)}_{\vec{k}}p^{(l)}_{\vec{k}}-H_{\phi}^{(l)}=\sum_{\vec{k} \in \mathcal{L}_+}\dot {\tilde{q}}^{(l)}_{\vec{k}}\tilde{p}^{(l)}_{\vec{k}}-\tilde{ H}_{\phi}^{(l)},
\end{align}
up to a total derivative, with the new Hamiltonian being given by
\begin{align}\label{eq:ham-realscamode}
\tilde{H}^{(l)}_{\phi}(N=a^3)  =&  \sum_{\vec{k} \in \mathcal{L}_+} \frac{b_{\tilde k}}{2} \left\{ \left(\tilde{p}^{(l)}_{\vec{k}}\right)^2 + \left[k^2 +s_{\tilde{k}}(\tau)\right] \left(\tilde{q}^{(l)}_{\vec{k}}\right)^2\right\},
\end{align}
where we have defined
\begin{align}\label{s}
s_{\tilde{k}}(\tau)=\frac{a^6m^{2}}{b_{\tilde k}^2}+\frac{3}{4}\left(\frac{\dot{b}_{\tilde k}}{b^2_{\tilde k}}\right)^2-\frac{1}{2}\frac{\ddot{b}_{\tilde k}}{b^3_{\tilde k}}.
\end{align}
We can see that $\tilde{H}^{(l)}_{\phi}(N=a^3)$ is a sum of Hamiltonians of harmonic oscillator type, one for each mode, up to the factor $b_{\tilde k}$ in Eq. \eqref{eq:ham-realscamode}. The mass of each of these oscillators depends on time, as well as on the unit wave vector $\vec{\tilde k}$. Note, nonetheless, that this mass $s_{\tilde{k}}$ does not depend on $k$, because $b_{\tilde k}$ is independent of this quantity.

\subsection{Classical dynamics}

Given the Hamiltonian \eqref{eq:ham-realscamode}, Hamilton's equations of motion for the different modes read
\begin{align}\label{eom}
\dot{\tilde{q}}_{\vec{k}}=b_{\tilde k} \tilde{p}_{\vec{k}}\qquad  \dot{\tilde{p}}_{\vec{k}}=-b_{\tilde k}(k^2+s_{\tilde{k}})\tilde{q}_{\vec{k}}.
\end{align}
Let us express their real solutions in the following general form:
\begin{eqnarray} \label{qeq}
\tilde q_{\vec{k}}(\tau) &=& Q_{\vec{k}} e^{i\Theta^{q}_{\vec{k}}(\tau)}+Q_{\vec{k}}^*e^{-i\left[\Theta^{q}_{\vec{k}}(\tau)\right]^*},\\ \label{peq}
\tilde p_{\vec{k}}(\tau) &=& k\left(P_{\vec{k}}e^{i\Theta^{p}_{\vec{k}}(\tau)}+P_{\vec{k}}^*e^{-i\left[\Theta^{p}_{\vec{k}}(\tau)\right]^*}\right), \label{p}
\end{eqnarray}
where $\Theta^{q}_{\vec{k}}(\tau)$ and $\Theta^{p}_{\vec{k}}(\tau)$ are complex functions, and the factor $k$ in Eq. \eqref{p} is introduced for later convenience. 

We do not need to determine exactly the complex solutions $e^{i\Theta^{q}_{\vec{k}}}$ and $e^{i\Theta^{p}_{\vec{k}}}$ for our study. Rather, it will be enough to know the asymptotic behavior of the functions $\Theta^{q}_{\vec{k}}(\tau)$ and $\Theta^{p}_{\vec{k}}(\tau)$ in the ultraviolet limit $k\rightarrow \infty$. This analysis is presented in Appendix \ref{appA}. With the initial conditions $\Theta^{x}_{\vec{k}}(\tau_0):= \Theta^{x}_{\vec{k},0}=0$ and $\dot\Theta^{x}_{\vec{k}}(\tau)|_{\tau_0}:=\Theta^{x}_{\vec{k},1}=b_{\tilde k}(\tau_0)k$ at an arbitrary initial harmonic time $\tau_0$, where  $x$ denotes either $q$ or $p$, we conclude that
\begin{align}\label{asymp}
\Theta^x_{\vec{k}}(\tau)&=k\eta_{\tilde k}(\tau)+\mathcal{O}(k^{-1}),
\end{align}
with
\begin{align}
\eta_{\tilde k}(\tau)=\int_{\tau_0}^{\tau}\,d\tilde\tau b_{\tilde k}(\tilde\tau).
\end{align}
The symbol $\mathcal{O}$ stands for {\it asymptotic order} in the ultraviolet regime. To obtain this result we have assumed that $\dot{s}_{\tilde k}$ exists and is integrable in every closed interval $[\tau_{0},\tau]$ of our time domain, where the function ${s}_{\tilde k}$ is defined in Eq. \eqref{s}.

Using at $\tau_0$ the equations of motion \eqref{eom}, as well as the definitions \eqref{qeq} and \eqref{peq}, we can express the constants $Q_{\vec{k}}$ and $P_{\vec{k}}$ in terms of the initial conditions $\tilde q_{\vec{k}}(\tau_0)$ and $\tilde p_{\vec{k}}(\tau_0)$, once the initial values $\Theta^{x}_{\vec{k},0}=0$ and $\Theta^{x}_{\vec{k},1}=b_{\tilde k}(\tau_0)k$ are given. Taking this into account, we can easily derive the expression of the modes $\tilde q_{\vec{k}}$ and $\tilde p_{\vec{k}}$, at any time $\tau$, in terms of  $\Theta^{q}_{\vec{k}}$, $\Theta^{p}_{\vec{k}}$, and those initial conditions. In matrix form
\begin{align}\label{evol}
&\begin{pmatrix}
\tilde q_{\vec{k}} \\ \tilde p_{\vec{k}} 
\end{pmatrix}_{\!\!\tau} =\mathcal{V}_{\vec{k}}(\tau,\tau_0)\begin{pmatrix}\tilde q_{\vec{k}} \\ \tilde p_{\vec{k}}  
\end{pmatrix}_{\!\!\tau_0},\quad \mathcal{V}_{\vec{k}}(\tau,\tau_0)=\begin{pmatrix}
\Re\left[e^{i\Theta^{q}_{\vec{k}}(\tau)}\right] & \frac{1}{k}\Im\left[e^{i\Theta^{q}_{\vec{k}}(\tau)}\right]  \\  
-\frac{k^2+s_{\tilde k}(\tau_0)}{k}\Im\left[e^{i\Theta^{p}_{\vec{k}}(\tau)}\right]  &
\Re\left[e^{i\Theta^{p}_{\vec{k}}(\tau)}\right]
\end{pmatrix}.
\end{align}
The subindex $\tau$ in column vectors means evaluation at that value of the harmonic time, and the symbols $\Re$ and $\Im$ denote the real and imaginary parts, respectively.

\section{Fock Quantization}\label{sec:fock}

After discussing the classical canonical description of the system, now we can proceed to analyze its Fock quantization. One can encode the freedom to select the Fock representation in the choice of what is called a complex structure $J$ \cite{wald}. The complex structure is a real linear transformation in the complex vector space $S$ such that $J^2=-1$. Moreover, 
$J$ should leave the product \eqref{inner} invariant, i.e. $\langle J\Phi_1 |J\Phi_2\rangle=\langle \Phi_1 |\Phi_2\rangle$. Every complex structure induces a splitting of $S$ into two mutually orthogonal subspaces, $S=S^+_J\oplus S^-_J$, where $S^\pm$ are the eigenspaces of $J$ with respective eigenvalues given by $\pm i$, namely $S^{\pm}=(S\mp iJS)/2$. In the subspace $S^+$ the product \eqref{inner} is a positive definite inner product. The usual convention is to assign the particle annihilationlike part of the solutions\footnote{For spacetimes with a timelike Killing vector and a natural concept of frequency, the subspace $S^+$ corresponds to the positive frequency solutions, while $S^-$ corresponds to the negative frequency ones.} to $S^+$. Its completion in the considered inner product is the one-particle Hilbert space, from which the symmetric Fock space is constructed. Therefore, different complex structures define different Fock spaces, corresponding to different annihilation and creation operators. In general, these different Fock representations are unitarily inequivalent and thus define physically distinct quantum theories. To remove this ambiguity and pick out a unique class of equivalent Fock representations, one can try and appeal to criteria based on well-founded physical motivations.

\subsection{Invariant Fock representations}

We will first make use of the symmetries of the system and focus our attention exclusively on Fock representations in which the vacuum is left invariant by the action of the group of Killing symmetries of the background corresponding to rigid rotations in $\mathbb{T}^3$ (obtained from the combination of individual rigid rotations in each of the diagonal directions). The complex structure of every such \emph{invariant} Fock representation commutes with the action of these symmetries, guaranteeing the vacuum invariance.  The most general form of a complex structure compatible with the considered background symmetries was in fact deduced in Refs. \cite{flat, flat2}. Equivalently, the invariant Fock representations can be characterized by specific properties of the corresponding choice of annihilationlike and creationlike variables (to later be declared annihilation and creation operators on Fock space), $a^{(l)}_{\vec k}$ and $a^{(l),\dagger}_{\vec k} := (a^{(l)}_{\vec k})^*$, related through a linear combination with the mode variables $(\tilde q^{(l)}_{\vec{k}},\tilde p^{(l)}_{\vec{k}})$. The admissible linear combinations turn out not to mix modes with different wave vectors $\vec{k}\in \cal{L}_+$ or different labels $l\in\{1,2\}$. A simple way to understand this result is the compatibility with the symmetries of the equations of motion in our background, which decouple these modes. On the other hand, the linear combinations that provide the annihilationlike and creationlike variables may depend on the wave vector $\vec{k}$. However, if we also respect the symmetry of the dynamical equations that makes them independent of the consideration of cosine or sine modes (corresponding to $l=1$ and 2, respectively), the combinations cannot depend on the label $l$ \cite{flat,flat2} (at the end of the day, this is so because the cosine and sine eigenfunctions of the Laplace-Beltrami operator are combinations of exponentials related by complex conjugation).

In summary, a symmetry-\emph{invariant} Fock representation is totally characterized by a linear transformation $\mathcal F$ formed by $2\times 2$ blocks $\mathcal{F}_{\vec{k}}$ such that 
\begin{align}\label{blockcs1}
\begin{pmatrix}
a^{(l)}_{\vec{k}}  \\ a^{(l),\dagger}_{\vec{k}} 
\end{pmatrix}=\mathcal{F}_{\vec{k}}
\begin{pmatrix}
\tilde q^{(l)}_{\vec{k}}  \\ \tilde p^{(l)}_{\vec{k}}
\end{pmatrix},\qquad \mathcal{F}_{\vec{k}}:=\begin{pmatrix}
f_{\vec{k}} & g_{\vec{k}} \\ f^*_{\vec{k}} & g^*_{\vec{k}}
\end{pmatrix}.
\end{align}
The coefficients $f_{\vec{k}}$ and $g_{\vec{k}}$ satisfy
\begin{align}\label{cond}
f_{\vec{k}}  g^*_{\vec{k}} -g_{\vec{k}}  f_{\vec{k}}^*=-i,
\end{align}
in order to guarantee that the annihilationlike and creationlike variables verify the characteristic Poisson relations
\begin{align}
 \{a^{(l)}_{\vec{k}},a^{({\tilde{l}}),\dagger}_{\vec{k}'}\}=-i\delta^{l,{\tilde{l}}}\delta_{\vec{k},\vec{k}'},
\end{align}
following from (\ref{eq:poisson-realmod}).

\subsection{Unitarity of the dynamics}

We restrict our discussion to invariant Fock representations from now on. We will consider families of representations connected by a nontrivial dynamics that is induced from the evolution of the Klein-Gordon field. Moreover, we want this dynamics to be implementable as a unitary operator. These families of representations will correspond to choices of annihilationlike and creationlike variables of the form \eqref{blockcs1}, but allowing the blocks $\mathcal{F}_{\vec{k}}$ that define those variables to smoothly depend on the evolution parameter, identified with the time $\tau$. This explicit time dependence will be able to absorb part of the time variation of the canonical field pairs $(\tilde q^{(l)}_{\vec{k}}, \tilde p^{(l)}_{\vec{k}})$, that evolve according to Eq. \eqref{evol}. In this way, we introduce a procedure to split the time variation of the Klein-Gordon field and its momentum into an explicit dependence on the time $\tau$, defined (at least locally) by the background average scale factor $a(\tau)$, and a dynamical evolution that is to be implemented quantum mechanically as a unitary transformation.

Before continuing our analysis, it is worth clarifying the role of the dynamics in the relation between the Fock representations that form each of the families that we are considering, representations that are determined by the sets of annihilationlike and creationlike variables  $a^{(l)}_{\vec{k}}(\tau)$ and $a^{(l),\dagger}_{\vec{k}}(\tau)$, where $\tau$ runs in our time domain. We can always take a given section of constant time, let us say at initial time $\tau_0$, and specify a Fock representation for our Klein-Gordon system by adopting as annihilationlike and creationlike variables the set formed by $a^{(l)}_{\vec{k}}(\tau_0)$ and $a^{(l),\dagger}_{\vec{k}}(\tau_0)$, regarded as coefficients in an expansion of the field in an appropriate basis of solutions. Alternatively, we can also express the variables $a^{(l)}_{\vec{k}}(\tau_0)$ and $a^{(l),\dagger}_{\vec{k}}(\tau_0)$ in terms of their evolved values $a^{(l)}_{\vec{k}}(\tau)$ and $a^{(l),\dagger}_{\vec{k}}(\tau)$ at any time $\tau$. The relation between these alternative sets will be given by a Bogoliubov transformation. The linearity of this transformation and of the space of solutions allows us to use these evolved values as new annihilationlike and creationlike coefficients on the {\emph{initial-time}} section. This new choice determines then a different Fock representation of the system, obtained from the previous one by the dynamical evolution of our annihilationlike and creationlike variables. The different representations defined in this way will be equivalent if and only if the introduced dynamics is implementable as a unitary operator on the Fock space associated with any of them (since they are ultimately equivalent). In the rest of this section we will discuss the necessary and sufficient conditions that a family of Fock representations has to fulfill in order to admit such a unitarily implementable dynamics. We will ignore again the superscript $l$ in our notation to simplify the formulas.

As we have anticipated, the classical evolution \eqref{evol} translates into a Bogoliubov transformation $\mathcal B$ that implements the dynamics of the annihilationlike and creationlike variables from the initial time $\tau_0$ to any time $\tau$:
\begin{align}\label{bog}
\begin{pmatrix} a_{\vec{k}} \\ a^{\dagger}_{\vec{k}} \end{pmatrix}_{\!\!\tau}=\mathcal{B}_{\vec{k}}(\tau,\tau_{0})\begin{pmatrix} a_{\vec{k}} \\ a^{\dagger}_{\vec{k}} \end{pmatrix}_{\!\!\tau_0},
\qquad \mathcal{B}_{\vec{k}}(\tau,\tau_{0}):=\begin{pmatrix} \alpha_{\vec{k}}(\tau,\tau_{0}) & \beta_{\vec{k}}(\tau,\tau_{0}) \\ \beta_{\vec{k}}^*(\tau,\tau_{0}) & \alpha_{\vec{k}}^*(\tau,\tau_{0}) \end{pmatrix}.
\end{align}
Recall that the subindex $\tau$ denotes evaluation at that instant of time. Using Eqs. \eqref{evol} and \eqref{blockcs1}, we can easily deduce the explicit form of this Bogoliubov transformation,
\begin{align}
\mathcal{B}_{\vec{k}}(\tau,\tau_{0})=\mathcal{F}_{\vec{k}}(\tau) \mathcal{V}_{\vec{k}}(\tau,\tau_0) \mathcal{F}^{-1}_{\vec{k}}(\tau_0).
\end{align}
Let us notice that this transformation is in fact independent of the label $l$, since we have seen that $\mathcal{F}_{\vec{k}}$ is $l$-independent for symmetry-invariant representations, and the evolution transformation $\mathcal{V}_{\vec{k}}$ is the same regardless of the value of $l$ [for more details about this statement, see Eq. \eqref{evol} and Appendix \ref{appA}]. 

The resulting alpha and beta coefficients are given by
\begin{align}
\alpha_{\vec{k}}(\tau,\tau_0)&=i\bigg\{ f_{\vec{k}}(\tau) g_{\vec{k}}^*(\tau_0) \Re[e^{i\Theta^q_{\vec{k}}(\tau)}]-g_{\vec{k}}(\tau) f_{\vec{k}}^*(\tau_0) \Re[e^{i\Theta^p_{\vec{k}}(\tau)}]
 \nonumber\\& -  \frac{1}{k} f_{\vec{k}}(\tau) f_{\vec{k}}^*(\tau_0)\Im[e^{i\Theta^q_{\vec{k}}(\tau)}]-\frac{k^2+s_{\tilde k}(\tau_0)}{k} g_{\vec{k}}(\tau) g_{\vec{k}}^*(\tau_0)\Im[e^{i\Theta^p_{\vec{k}}(\tau)}]\bigg\},\\
\beta_{\vec{k}}(\tau,\tau_0)&=-i\bigg\{ f_{\vec{k}}(\tau) g_{\vec{k}}(\tau_0) \Re[e^{i\Theta^q_{\vec{k}}(\tau)}]-g_{\vec{k}}(\tau) f_{\vec{k}}(\tau_0) \Re[e^{i\Theta^p_{\vec{k}}(\tau)}]
 \nonumber\\& - \frac{1}{k} f_{\vec{k}}(\tau) f_{\vec{k}}(\tau_0)\Im[e^{i\Theta^q_{\vec{k}}(\tau)}]-\frac{k^2+s_{\tilde k}(\tau_0)}{k} g_{\vec{k}}(\tau) g_{\vec{k}}(\tau_0)\Im[e^{i\Theta^p_{\vec{k}}(\tau)}]\bigg\}.\label{beta}
\end{align}

It is well known that such a Bogoliubov transformation is implementable as a unitary operator on Fock space if and only if its antilinear part is Hilbert-Schmidt \cite{Shale}, which in turn is true if and only if
\begin{align}\label{ucond}
\sum_{\vec{k}\in\cal L_+}|\beta_{\vec{k}}(\tau,\tau_0)|^{2}<\infty,
\end{align}
for all times $\tau$ in the evolution. This last condition can be equivalently interpreted as the requirement that the evolved vacuum, annihilated by the operators representing all the variables $a^{(l)}_{\vec{k}}(\tau)$ at time $\tau$, have a finite particle content from the point of view of the vacuum defined at the initial time $\tau_0$. 

To prove under which conditions this is verified, we need the behavior of $|\beta_{\vec{k}}|$ for asymptotically large $k$, i.e., in the limit $k\rightarrow \infty$. Taking into account Eq. \eqref{asymp}, we obtain
\begin{align}\label{norm-beta}
|\beta_{\vec{k}}(\tau,\tau_0)|&=\frac{1}{2k}\Big| \left[f_{\vec{k}}(\tau)+ikg_{\vec{k}}(\tau)\right]\left[f_{\vec{k}}(\tau_0)-ikg_{\vec{k}}(\tau_0)
\right] \left[e^{ik\eta_{\tilde k}(\tau)}+\mathcal{O}(k^{-1})\right] \nonumber
\\& -\left[f_{\vec{k}}(\tau)-ikg_{\vec{k}}(\tau)\right]\left[f_{\vec{k}}(\tau_0)+ikg_{\vec{k}}(\tau_0)\right] \left[e^{-ik\eta_{\tilde k}(\tau)}+\mathcal{O}(k^{-1})\right] \nonumber
\\&
+2ig_{\vec{k}}(\tau)g_{\vec{k}}(\tau_0)s_{\tilde k}(\tau_0)\left\{\sin{[k\eta_{\tilde k}(\tau)]}+\mathcal{O}(k^{-1})\right\}\Big|. 
\end{align}
Notice that the last term in this formula is subdominant with respect to the terms proportional to $g_{\vec{k}}(\tau)g_{\vec{k}}(\tau_0)$ in the second and first terms.

Given the above expression of the beta coefficients, the unitarity condition \eqref{ucond} imposes a specific behavior of the functions $f_{\vec{k}}(\tau)$ and $g_{\vec{k}}(\tau)$ in the limit $k\rightarrow\infty$. If one recalls the relation \eqref{cond} between these functions, the condition restricts severely the asymptotic form of, e.g., $f_{\vec{k}}(\tau)$. We discard the possibility of compensating the contribution coming from the first term of \eqref{norm-beta} with the contribution coming from the second term in order to arrive at a square summable sequence of beta coefficients. This would imply a trivialization of the dynamics, absorbing the dominant dynamical contribution of the Fourier modes of our canonical variables by tuning in a specific and counterbalanced way the time dependence of the functions $f_{\vec{k}}$ and $g_{\vec{k}}$. Let us then look at these two terms of \eqref{norm-beta} separately, admitting their functional independence as functions of time. 

It is not hard to see that the only possibility to attain Eq. \eqref{ucond} in a consistent way with restriction \eqref{cond} is to require $ f_{\vec{k}}=-ikg_{\vec{k}}$ for all $\vec{k}\in\cal L_+$, except perhaps a finite number of wave vectors, and up to negligible terms in the ultraviolet regime. This condition, together with Eq. \eqref{cond}, fixes completely the leading term (up to a phase) of both $f_{\vec{k}}$ and $g_{\vec{k}}$ in the asymptotic expansion for large $k$. Indeed, for all $\vec{k}\in\cal L_+^\prime$, such that $\cal L_+^\prime$ differs from $\cal L_+$ in a finite number of elements at most, we need to have
\begin{align}\label{fg-asym}
f_{\vec{k}}(\tau)=\sqrt{\frac{k}{2}}e^{iF_{\vec{k}}(\tau)}+k\theta^{f}_{\vec{k}}(\tau),\qquad 
g_{\vec{k}}(\tau)=\frac{i}{\sqrt{2k}}e^{iF_{\vec{k}}(\tau)}+\theta^{g}_{\vec{k}}(\tau),
\end{align}
where $F_{\vec{k}}$ is any phase, and $k\theta^{f}_{\vec{k}}(\tau)$ and $\theta^{g}_{\vec{k}}(\tau)$ are subdominant terms in the asymptotic limit of infinite $k$. Furthermore, these subdominant contributions must satisfy
\begin{align}\label{ucond2}
\sum_{\vec{k}\in{\cal L_+^\prime}}k|\theta^{f}_{\vec{k}}(\tau)+i\theta^{g}_{\vec{k}}(\tau)|^{2}<\infty,
\end{align}
for all $\tau$. To arrive at this conclusion, we have used the fact that the sequence $\{s_{\tilde k}(\tau_0)/k^{2}\}_{\vec{k}\in\cal L_+}$ is square summable. The function $s_{\tilde k}$ does not depend on $k$ and the sequence $\{1/k^{2}\}_{\vec{k}\in\cal L_+}$ is square summable, as proven in Ref. \cite{flat}.

It is worth recalling that the functions $f_{\vec{k}}(\tau)$ and $g_{\vec{k}}(\tau)$ are independent of the value of the label $l$, and therefore the same applies to the functions appearing in their asymptotic expansions.  On the other hand, we note that, owing to relation \eqref{cond}, the subdominant terms are not independent, but they satisfy the relation
\begin{align}\label{rel}
\Re\left[i\left(e^{-iF_{\vec{k}}}+\sqrt{2k}(\theta^{f}_{\vec{k}})^*\right)\theta^{g}_{\vec{k}}\right]=\Re\left[\theta^{f}_{\vec{k}}e^{-iF_{\vec{k}}}\right].
\end{align}
Calling $\theta^{g}_{\vec{k}}=|\theta^{g}_{\vec{k}}|e^{iG_{\vec{k}}}$, this equation gives the expression of $|\theta^{g}_{\vec{k}}|$ (at least for sufficiently large $k$) in terms of $F_{\vec{k}}$, $\theta^{f}_{\vec{k}}$, and the phase $G_{\vec{k}}$, unless the latter is made to coincide with $F_{\vec{k}}$ (modulo $2\pi$) up to negligible terms when $k\rightarrow \infty$. In all other cases, one can employ the expression obtained for $|\theta^{g}_{\vec{k}}|$ and demonstrate that condition \eqref{ucond2} is satisfied if and only if the sequence $\{\sqrt{k}|\theta^{f}_{\vec{k}}|\}_{\vec k \in\cal L_+^\prime}$ is square summable.

One can  convince oneself that the situation where $f_{\vec{k}}(\tau)$ and $g_{\vec{k}}(\tau)$ are time independent can be addressed along the lines of our general discussion and that one arrives then to similar conclusions.\footnote{For details on an alternate proof 
concerning the time-independent situation see e.g. Ref. \cite{cmv2}.}  In this case, the square summability of the beta coefficients, together with relation \eqref{cond}, again implies that $ f_{\vec{k}}=-ikg_{\vec{k}}$, up to terms that are negligible. One obtains again the behavior shown in Eq. \eqref{fg-asym}, particularized to phases $F_{\vec{k}}$ and subdominant contributions $\theta^{f}_{\vec{k}}$ and $\theta^{g}_{\vec{k}}$ that are constant in time.

In conclusion, expressions \eqref{fg-asym}, \eqref{ucond2}, and \eqref{rel} are the necessary and sufficient conditions that a time-dependent family of invariant Fock representations, characterized by $\mathcal{F}(\tau)$, has to verify in order that its elements are related by a dynamical evolution that can be implemented as a unitary endomorphism in Fock space. We would like to comment that, thanks to the canonical transformation carried out in Sec. \ref{trans}, the norms of the leading terms of $f_{\vec{k}}$ and $g_{\vec{k}}$ are time independent (and therefore background independent), as well as mass independent. The time and mass dependence may appear only in the phase $F_{\vec k}$ of these leading terms, and in the subleading ones. In particular, this implies that the condition of unitary implementability is capable of fixing the dynamics of the annihilationlike and creationlike variables at dominant order, up to an irrelevant phase, once one has imposed the invariance of the vacuum under the background symmetries.

To conclude this section, let us emphasize that the analysis can be carried out equally well without performing the canonical time-dependent transformation \eqref{canonical}, expressing the families of annihilationlike and creationlike variables corresponding to invariant representations directly in terms of the original canonical field variables $q_{\vec{k}}$ and $p_{\vec{k}}$. Obviously, the results attained in that way and those presented here are the same, as we discuss in Appendix \ref{appB}.

\subsection{Massless representation}

Among all possible Fock representations verifying conditions \eqref{fg-asym}-\eqref{rel}, there is an especially simple one, which is given by
\begin{align}\label{massless}
\breve f_{\vec{k}}(\eta)=\sqrt{\frac{k}{2}},\qquad 
\breve g_{\vec{k}}(\eta)=\frac{i}{\sqrt{2k}}.
\end{align}
As one can check by simple inspection of Eq. \eqref{norm-beta}, in this representation we get that  $|\beta_{\vec{k}}|=|s_{\tilde k}(\tau_0)\mathcal{O}(k^{-2})|$. It is the natural representation that one would adopt in a static spacetime for a scalar field with configuration and momentum variables given by ${\tilde{q}}_{\vec{k}}$ and ${\tilde{p}}_{\vec{k}}$ if the time-dependent mass term $s_{\tilde{k}}$ were zero. We will employ this representation as a \emph{reference} representation in the next section. We denote by $\breve a_{\vec{k}}$ and $\breve{a}^{\dagger}_{\vec{k}}$ the associated annihilationlike and creationlike variables.

\section{Uniqueness of the Fock representation}
\label{sec:uni}

We have characterized the Fock quantizations possessing vacua that are invariant under the symmetries of the Bianchi I background, and a quantum dynamics that can be implemented unitarily. In this section, we will prove that it is not only that the representations connected dynamically in each of these quantizations are unitarily equivalent (given the unitary implementability of the quantum dynamics), but, furthermore, the representations of any of these possible quantizations are all equivalent between them. In particular, they are unitarily equivalent to our reference representation at any value of the time parameter. As a consequence of this result, the criterion that we are putting forward indeed guarantees the uniqueness of the choice of Fock representation for the scalar field, up to unitary equivalence, in the considered cosmologies. 

Let us consider the \emph{reference} Fock representation, and any other representation with annihilationlike and creationlike variables $(a_{\vec{k}},a^{\dagger}_{\vec{k}})$ selected by our criterion. These annihilationlike and creationlike variables at generic time $\tau$ are related to the reference ones, $(\breve a_{\vec{k}},\breve a^{\dagger}_{\vec{k}})$, by means of a Bogoliubov transformation:
\begin{align}\label{bogK}
\begin{pmatrix}  a_{\vec{k}} \\ a^{\dagger}_{\vec{k}} \end{pmatrix}_{\!\!\tau}=\mathcal{K}_{\vec{k}}(\tau)\begin{pmatrix} \breve a_{\vec{k}} \\ \breve{a}^{\dagger}_{\vec{k}} \end{pmatrix}_{\!\!\tau},\qquad  \mathcal{K}_{\vec{k}}:=\begin{pmatrix} \kappa_{\vec{k}} & \lambda_{\vec{k}} \\ \lambda^*_{\vec{k}} & \kappa^*_{\vec{k}} \end{pmatrix}.
\end{align}
Note that, in general, $\mathcal{K}_{\vec{k}}$ can be time dependent. 

It is straightforward to show that $\mathcal{K}_{\vec{k}}={\mathcal{F}}_{\vec{k}}(\breve{\mathcal{F}}_{\vec{k}})^{-1}$ (and hence it is independent of the value of the label $l$). Using this relation, we obtain 
\begin{align}\label{k}
\kappa_{\vec{k}}(\tau)&=i\left[ f_{\vec{k}}(\tau)\breve{g}^*_{\vec{k}}(\tau)-g_{\vec{k}}(\tau)\breve{f}^*_{\vec{k}}(\tau)\right],\\
\lambda_{\vec{k}}(\tau)&=-i\left[ f_{\vec{k}}(\tau)\breve{g}_{\vec{k}}(\tau)-g_{\vec{k}}(\tau)\breve{f}_{\vec{k}}(\tau)\right].\label{l}
\end{align}

The two Fock representations considered at (an arbitrary value of the) time $\tau$ are unitarily equivalent if and only if the above Bogoliubov transformation, formed by the sequence of matrices $\mathcal{K}_{\vec{k}}$, can be implemented as a unitary operator in the quantum theory, namely if and only if its antilinear part is Hilbert-Schmidt \cite{Shale}. This in turn amounts to demand that
\begin{align}\label{un}
\sum_{\vec{k}\in\cal L_+}|\lambda_{\vec{k}}(\tau)|^{2}<\infty.
\end{align}
Taking into account the asymptotic behavior of  $f_{\vec{k}}$ and $g_{\vec{k}}$, given in Eq. \eqref{fg-asym}, we get from expression \eqref{l} that
\begin{align}
\sum_{\vec{k}\in{\cal L_+^\prime}}|\lambda_{\vec{k}}(\tau)|^{2}=\frac12\sum_{\vec{k}\in{\cal L_+^\prime}}k|\theta^f_{\vec{k}}(\tau)+i\theta^g_{\vec{k}}(\tau)|^{2}.
\end{align}
The finiteness of this sum is just the necessary and sufficient condition \eqref{ucond2} for a unitary implementability of the dynamics, implementability that we are assuming for the quantization determined by $(a_{\vec{k}}(\tau),a^{\dagger}_{\vec{k}}(\tau))$. Therefore, the considered quantizations are unitarily equivalent, as we wanted to show.

In conclusion, the criterion of imposing symmetry invariance and unitarity of the dynamics picks out a unique equivalence class of invariant Fock representations.

\section{Conclusions}\label{sec:concl}

In this paper we have addressed the question of the uniqueness of the Fock quantization of a minimally coupled test scalar field with mass $m$ in a Bianchi I spacetime with compact spatial slices. Extending to this anisotropic situation previous results obtained in flat isotropic cosmologies \cite{flat,flat2}, we have proven that the requirements of (i) invariance of the vacuum under spatial Killing symmetries, and (ii) a quantum dynamics that can be implemented as a unitary endomorphism in Fock space, select a unique Fock representation, up to unitary equivalence, removing the ambiguity inherent to quantum field theory in curved backgrounds. Moreover, our requirements determine as well which part of the field evolution must be dealt with as a background (or, equivalently, explicit time) dependence and which part can be promoted to a quantum dynamics with the desired properties.

We started by analyzing the field dynamics of the classical system. We first carried out a mode decomposition of the canonical field and its momentum in terms of the eigenmodes of the Laplace-Beltrami operator defined on the spatial sections. Its eigenspaces are spanned by plane waves labeled with a wave vector of integers $\vec{k}$ and a dichotomic variable $l$, corresponding either to cosine or to sine modes. Afterwards, we introduced a canonical transformation of the field variables that does not mix different modes (in consonance with the realization that they are not dynamically coupled). Most importantly, this canonical transformation is time dependent, so that it extracts part of the evolution of the original variables, rendering the dynamics of the redefined field variables simpler. Indeed, the resulting Hamiltonian is a sum of terms, one for each mode, that, up to a mode-dependent global factor, resemble Hamiltonians of harmonic oscillators of wave number $k$ equal to the norm of the corresponding wave vector, but in the presence of a time-dependent potential that is different for each mode. Then, we were able to solve the resulting classical dynamics to leading and first subleading orders in an asymptotic expansion for infinitely large $k$. Moreover, the introduced canonical transformation is actually nonlocal, in the sense that the  change of configuration and momentum variables is mode dependent, and hence depends on global features of the Laplace-Beltrami operator. In the case of the scalar field that we are considering, this nonlocality can be traced back to the fact that the spatial sections of the cosmology are anisotropic. In comparison, in isotropic situations, this canonical transformation reduces to a local one, defined independently at each spacetime point (inasmuch as we can regard the metric components as local spacetime functions). In particular, for the configuration field variable this local transformation amounts to a rescaling by the scale factor of the cosmology \cite{flat,flat2}.

We then proceeded to discuss the issue of the choice of a Fock representation for the quantization of the field.
We have focused our attention on representations--or associated complex structures--that are invariant under the spatial Killing symmetries of the background and treat on an equal footing the cosine and sine modes, since the interchange of these modes is a symmetry of the dynamics. The complex structures that are invariant under these symmetries can be characterized by two coefficients for each wave vector $\vec{k}$. These coefficients, $f_{\vec{k}}$ and $g_{\vec{k}}$, define for each mode the linear combinations of configuration and momentum variables that determine the annihilationlike and creationlike variables of the representation.

Among these types of Fock representations, our interest has been focused on time-dependent families, described by two sequences of functions  $f_{\vec{k}}(\tau)$ and $g_{\vec{k}}(\tau)$, such that they can be connected by means of a nontrivial dynamics that is unitarily implementable. Together with the background dependence of the field and its momentum, this unitary dynamics has to reproduce the evolution of the Klein-Gordon system. The requirement of a nontrivial unitary implementation turns out to constrain not only this dynamics, but also the asymptotic behavior of the functions $f_{\vec{k}}(\tau)$ and $g_{\vec{k}}(\tau)$ when $k\to\infty$. Indeed, we have shown that, in each of these functions, the dominant term in an asymptotic expansion for infinitely large $k$ is completely fixed. Thanks to the mode-dependent canonical transformation that we performed on the field configuration and momentum variables, these dominant contributions have norms that are mass and time independent, and depend on the wave vector $\vec{k}$ only through the wave number $k$ [see Eq. \eqref{fg-asym}]. Actually, the canonical transformation extracts precisely the part of the Klein-Gordon dynamics that cannot be implemented unitarily in the quantum theory. The remaining dynamics can be promoted to a unitary quantum mechanical transformation, and it is fixed up to subdominant contributions that, in any case, respect the unitary equivalence of the Fock representations related by the evolution. 

Finally, we have demonstrated that the different time-dependent families of invariant Fock representations selected by our criterion of symmetry invariance and unitary dynamics are in fact all equivalent. Therefore, our criterion fixes a unique Fock representation (up to unitary equivalence) and essentially a unique nontrivial quantum dynamics (in the sense explained above). One particularly simple representative within the class of admissible Fock representations is the  representation determined by relations \eqref{massless}, that one would naturally choose for a massless field in a static spacetime.

It is worth emphasizing that our proof of uniqueness does not need the time parameter to be defined on the whole real line, not even in an unbounded subset of it. It suffices that there exists a connected interval where our arguments can be applied. In particular, this is so for the definition of the function $b_{\tilde k}$ in Eq. \eqref{bk} (which might blow up outside of the considered interval, if a directional scale factor vanished there) and for the smoothness requirements on the function $s_{\tilde k}(\tau)$, that must have a derivative that is integrable in any closed subinterval of the considered time domain (see Appendix \ref{appA}).

This is the first example where our uniqueness criterion, which has been typically studied in the context of cosmological scenarios, is applied to a scalar field in a background that is not conformally ultrastatic. Indeed, this particular example provides insights about how a quantization compatible with spatial symmetries and a unitary dynamics can be extended to quantum field theories in general homogeneous spacetimes, that can be anisotropic. Another interesting example of anisotropic cosmologies is the case of Bianchi III spacetimes. For some specific backgrounds of that type that are shear free, the cosmological perturbations turn out to fulfill equations of motion that, at least in certain regimes, are similar to those of a Klein-Gordon field in a static isotropic spacetime with a time-dependent potential, as shown in Ref. \cite{cmp}. Therefore, in those cases and in such regimes, we expect that one may define a local (i.e., mode-independent) canonical transformation capable of extracting the part of the field dynamics that is not unitarily implementable in the quantum theory, like in the isotropic case. 

As we mentioned in the Introduction, our results are potentially applicable to the quantization of cosmological perturbations in Bianchi I spacetimes. In this case, the coupling between tensor and scalar modes introduces subtleties that must be carefully analyzed, but a preliminary study suggests that the coupling does not affect the relevant dominant terms of the ultraviolet asymptotic expansion of the physical perturbations, supporting the expectation that our uniqueness results can be adapted to this framework. In addition, these uniqueness results can also be applied in the context of quantum cosmology, where the background is not treated as a classical spacetime but quantum mechanically. One of the most interesting formalisms in these lines is loop quantum cosmology \cite{revLQC1,revLQC2}. In the case of Bianchi I cosmologies, although their quantization is not completely understood within this formalism yet, it is nonetheless possible to compute effective geometries. Then, one can consider quantum field theories in those effective spacetimes. The uniqueness criterion put forward here is expected to be valid in those situations as well. 
 
\acknowledgments
This work was partially supported by the research grant MINECO Project No. FIS2014-54800-C2-2-P from Spain, COST Action MP1405 QSPACE, supported by  European Cooperation in Science and Technology (COST), and DGAPA-UNAM IN113115 and CONACyT 237351 from Mexico. In addition, M. M.-B. acknowledges financial support from the Netherlands Organisation for Scientific Research (NWO), and J.O. from NSF grants NSF-PHY-1305000, PHY-1505411, the Eberly research funds of Penn State (USA) and Pedeciba (Uruguay).

\appendix

\section{Asymptotic analysis of the dynamics}
\label{appA}

In this appendix, we study the asymptotic behavior for large wave vector $k$ of the complex solutions $e^{i\Theta^{q}_{\vec{k}}}$ and $ke^{i\Theta^{p}_{\vec{k}}}$ to the equations of motion \eqref{eom} for the canonical variables of the Klein-Gordon field, $q_{\vec{k}}$ and $p_{\vec{k}}$, respectively. Obviously, the complex conjugate functions $e^{-i(\Theta^{q}_{\vec{k}})^*}$ and $ke^{-i(\Theta^{p}_{\vec{k}})^*}$ are also solutions to the considered dynamical equations, because those differential equations have coefficients that are given by real functions of time. Using this fact, we can choose freely the sign of the real part of $\Theta^{q}_{\vec{k}}$ and $\Theta^{p}_{\vec{k}}$ in the following discussion.

\subsection{Asymptotic analysis of $\tilde q_{\vec{k}}$}

Let us first focus on the function $\Theta^{q}_{\vec{k}}$ that determines the solution for $\tilde q_{\vec{k}}$.  We follow the analysis of Ref. \cite{cmv2}. For those readers familiar with this analysis, it is enough to realize that the choice of time $d\eta_{\tilde k} =b_{\tilde k}d\tau $ formally allows us to follow without further considerations the calculations of Ref. \cite{cmv2}, keeping in mind that at the end we should write all the expressions in terms of the original time $\tau$ (or any other choice of global time, common to all modes). Note that $d\eta_{\tilde k}$ is proportional to the conformal time, with a proportionality factor that, strictly speaking, would depend on the (unit) directional wave vector of the mode, though not on its wave number. Here, we will show the calculations explicitly for the time $\tau$.

Combining Hamilton's equations of motion for $(\tilde q_{\vec{k}},\tilde p_{\vec{k}})$, we obtain the following second-order, ordinary differential equation for $\tilde q_{\vec{k}}$:
\begin{align}\label{eq:tildq-eq}
&\frac{1}{b_{\tilde k}}\frac{d}{d\tau}\left(\frac{1}{b_{\tilde k}}\frac{d}{d\tau}\tilde q_{\vec{k}}\right)+\left(k^2 +s_{\tilde k}\right) \tilde q_{\vec{k}}=0.
\end{align}
Let us look for the complex solution $e^{i\Theta^{q}_{\vec{k}}}$ with the following expression for the phase $\Theta^{q}_{\vec{k}}$:
\begin{align}\label{A2}
\dot\Theta^q_{\vec{k}}(\tau)=b_{\tilde k}(\tau)\left[k+iW^q_{\vec{k}}(\tau)\right],
\end{align}
so that, taking the initial condition $\Theta^{q}_{\vec{k},0}=0$, we have
\begin{align}\label{eq:thta-vs-w}
\Theta^q_{\vec{k}}(\tau)=\int_{\tau_0}^\tau d\tilde\tau\, b_{\tilde k}(\tilde\tau)\left[k+iW^q_{\vec{k}}(\tilde\tau)\right].
\end{align}
If we now substitute this expression into Eq. \eqref{eq:tildq-eq}, we can see that $W^q_k(\tau)$ has to satisfy the first-order differential equation
\begin{align}\label{eq:w-eq}
\frac{1}{b_{\tilde k}}\dot W^q_{\vec{k}} = (W^q_{\vec{k}})^2-2ik  W^q_{\vec{k}}+ s_{\tilde k},
\end{align}
Let us now assume that $W^q_{\vec{k}}=\mathcal{O}(k^{-1})$ in the ultraviolet regime $k\rightarrow\infty$, so that it tends to the solution  $\tilde W^q_{\vec{k}}$ of
\begin{align}
\frac{1}{b_{\tilde k}}\dot{\tilde W}^q_{\vec{k}} = -2ik  
\tilde W_{\vec{k}}^q +s_{\tilde k}.
\end{align}
This equation, with initial condition $\tilde W^q_{\vec{k}}(\tau_0)=0$, is solved by
\begin{align}
\tilde W^q_{\vec{k}}(\tau)=e^{-2ik\eta_{\tilde k}(\tau)}\int_{\tau_0}^\tau\,d\tilde\tau b_{\tilde k}(\tilde\tau)s_{\tilde k}(\tilde\tau)e^{2ik\eta_{\tilde k}(\tilde\tau)}.
\end{align}
We recall that
\begin{align}
\eta_{\tilde k}(\tau)=\int_{\tau_0}^{\tau}\,d\tilde\tau b_{\tilde k}(\tilde\tau).
\end{align}
Integrating by parts, we obtain 
\begin{align}\nonumber
\tilde W^q_{\vec{k}}(\tau)=&-\frac{i}{2k}\left\{s_{\tilde k}(\tau)-s_{\tilde k}(\tau_0)e^{-2ik\eta_{\tilde k}(\tau)}-e^{-2ik\eta_{\tilde k}(\tau)}\int_{\tau_0}^\tau\,d\tilde\tau \dot{s}_{\tilde k}(\tilde\tau)e^{2ik\eta_{\tilde k}(\tilde\tau)}\right\}.
\end{align}
Since $s_{\tilde k}$ does not depend on the wave number $k$, then, if $ \dot{s}_{\tilde k}$ exists and is integrable in every closed interval $[\tau_{0},\tau]$ of the considered time domain, there exists a non-negative function $C(\tau)$ that is $k$-independent 
(but may depend on the unit wave vector) and such that 
\begin{align}
|\tilde W^q_{\vec{k}}(\tau)|\leq \frac{C(\tau)}{k}.
\end{align}
We see then that the function $\tilde W^q_{\vec{k}}(\tau)$ behaves as $\mathcal{O}(k^{-1})$ in the ultraviolet limit of large $k$ and, consistently with our assumptions, can be taken in fact as an asymptotic solution of Eq. \eqref{eq:w-eq}, up to higher-order corrections. Plugging this result into Eq. \eqref{eq:thta-vs-w}, we conclude that, in the ultraviolet regime,
\begin{align}
\Theta^q_{\vec{k}}(\tau)=k\eta_{\tilde k}(\tau)+\mathcal{O}(k^{-1}).
\end{align}

\subsection{Asymptotic analysis of $\tilde p_{\vec{k}}$}

To obtain the solution $ke^{i\Theta^{p}_{\vec{k}}}$ for $\tilde p_{\vec{k}}$, we use the first identity in Eq. \eqref{eom}, which relates $\Theta^{p}_{\vec{k}}$ to a complex solution of the equation of motion for $\tilde q_{\vec{k}}$. Expressing the latter as a linear combination of $e^{i\Theta^{q}_{\vec{k}}}$ and its complex conjugate, with respective coefficients called $A_{\vec{k}}$ and $B_{\vec{k}}$,
we get, employing also definition \eqref{A2}:
\begin{equation}\label{thetapq}
e^{i\Theta^{p}_{\vec{k}}}= A_{\vec{k}} \left[i-\frac{W^q_{\vec{k}}(\tau)}{k}\right]
e^{i\Theta^{q}_{\vec{k}}}- B_{\vec{k}} \left[i+\frac{(W^{q}_{\vec{k}})^*(\tau)}{k}\right]
e^{-i(\Theta^{q}_{\vec{k}})^*},
\end{equation}
with $\Theta^{q}_{\vec{k}}$ given in terms of $W^q_{\vec{k}}$ in Eq. \eqref{eq:thta-vs-w}. We choose initial conditions such that $\Theta^p_{\vec{k},0}=0$ and $\Theta^p_{\vec{k},1}=b_{\tilde k}(\tau_0)k$. Recalling that $W^q_{\vec{k}}(\tau_0)=0$, and realizing that Eq. \eqref{eq:w-eq} implies then that 
\begin{equation}{\dot W}^q_{\vec{k}}(\tau_0)=b_{\tilde k}(\tau_0)s_{\tilde k}(\tau_0),
\end{equation} our initial conditions select the following values for the coefficients $A_{\vec{k}}$ and $B_{\vec{k}}$:
\begin{align}\label{AB}
A_{\vec{k}}=& -\frac{i}{2} \frac{2 k^2 + s_{\tilde k}(\tau_0)}{ k^2 + s_{\tilde k}(\tau_0)}=-i+\mathcal{O}(k^{-2}),\\
\label{BA} B_{\vec{k}}=& \frac{i}{2}\frac{s_{\tilde k}(\tau_0)}{ k^2 + s_{\tilde k}(\tau_0)}=\mathcal{O}(k^{-2}) .
\end{align}

Finally, taking into account that $W^q_{\vec{k}}(\tau)=\mathcal{O}(k^{-1})$, as we have seen above, and that $b_{\tilde k}$ and $s_{\tilde k}$ are independent of the wave number $k$, it is straightforward to conclude from the above formulas that 
\begin{align}
\Theta^p_{\vec{k}}(\tau)=k\eta_{\tilde k}(\tau)+\mathcal{O}(k^{-1}).
\end{align}

\section{Back to the original description of the field}
\label{appB} 

In Sec. \ref{trans} we introduced the time-dependent canonical transformation \eqref{canonical} before performing our analysis of the unitarity of the dynamics. The reason is that the transformed modes $(\tilde q_{\vec{k}}, \tilde p_{\vec{k}})$ verify a simpler dynamics than the original ones $(q_{\vec{k}}, p_{\vec{k}})$, and as a result the analysis of the involved Bogoliubov transformations is much simpler. Nevertheless, we could have carried out the whole discussion without introducing such canonical transformation, defining directly the annihilationlike and creationlike variables $(a_{\vec{k}},a^{\dagger}_{\vec{k}})$ by means of a (possibly) time-dependent linear transformation $\tilde{\mathcal F}(\tau)$ of the original modes $(q_{\vec{k}}, p_{\vec{k}})$, namely 
\begin{align}\label{blockcs1bis}
\begin{pmatrix}
a_{\vec{k}}  \\ a^{\dagger}_{\vec{k}} 
\end{pmatrix}_{\!\!\tau}=\tilde{\mathcal{F}}_{\vec{k}}(\tau)
\begin{pmatrix}
 q_{\vec{k}}  \\  p_{\vec{k}}
\end{pmatrix}_{\!\!\tau},\qquad \tilde{\mathcal{F}}_{\vec{k}}(\tau)=\begin{pmatrix}
\tilde f_{\vec{k}}(\tau) & \tilde g_{\vec{k}}(\tau) \\ \tilde f^*_{\vec{k}}(\tau) & \tilde g^*_{\vec{k}}(\tau)
\end{pmatrix},
\end{align}
with
\begin{align}\label{condbis}
\tilde f_{\vec{k}}  \tilde g^*_{\vec{k}} -\tilde g_{\vec{k}}  \tilde f_{\vec{k}}^*=-i
\end{align}
at all times. 

Obviously, the time variation of the Klein-Gordon variables $(q_{\vec{k}}, p_{\vec{k}})$ extracted by the linear transformation $\tilde{\mathcal{F}}_{\vec{k}}(\tau)$ includes the part that was already extracted by the canonical transformation \eqref{canonical}, since $\tilde{\mathcal{F}}_{\vec{k}}(\tau)={\mathcal{F}}_{\vec{k}}(\tau){\mathcal{P}}_{\vec{k}}(\tau)$, where we have used that our change of canonical variables consists just in a linear transformation ${\mathcal{P}}_{\vec{k}}(\tau)$. Employing this identity for $\tilde{\mathcal{F}}_{\vec{k}}(\tau)$, we get 
\begin{align}
 \tilde f_{\vec{k}}=\sqrt{b_{\tilde k}} f_{\vec{k}}+\frac12\frac{\dot{b}_{\tilde k}}{b_{\tilde k}\sqrt{b_{\tilde k}}}  g_{\vec{k}},\qquad  \tilde g_{\vec{k}}=\frac1{\sqrt{b_{\tilde k}}} g_{\vec{k}}.
\end{align}

Clearly, had we done the analysis using the characterization of the Fock representation via Eq. \eqref{blockcs1bis}, we would have concluded that the necessary and sufficient conditions for the dynamics to admit a unitary implementation is that the functions $\tilde f_{\vec{k}}$ and $\tilde g_{\vec{k}}$ have the following asymptotic expansion in the ultraviolet regime of infinitely large $k$:
\begin{align}\label{fgbis}
\tilde f_{\vec{k}}=\sqrt{b_{\tilde k}} \left[\sqrt{\frac{k}{2} } \left(1+i\frac{\dot{b}_{\tilde k}} {2kb_{\tilde k}^2} \right) e^{iF_{\vec{k}}} + k  \tilde\theta^{f}_{\vec{k}} \right] ,\qquad 
\tilde{g}_{\vec{k}}=\frac{1} {\sqrt{b_{\tilde k}}} \left( \frac{i}{\sqrt{2k}} e^{iF_{\vec{k}}}+\tilde\theta^{g}_{\vec{k}} \right) , 
\end{align}
where $F_{\vec{k}}$ is any phase, while the subdominant terms $\tilde\theta^{f}_{\vec{k}}(\tau)$ and $\tilde\theta^{g}_{\vec{k}}(\tau)$ are related through Eq. \eqref{condbis} and must satisfy 
\begin{align}
\sum_{\vec{k}\in\cal L_+^\prime} k \left\vert \tilde\theta^{f}_{\vec{k}}+\left(i -\frac{\dot{b}_{\tilde k}(\tau)}{2kb_{\tilde k}^2(\tau)}\right)\tilde\theta^{g}_{\vec{k}}(\tau)\right\vert^{2}<\infty.
\end{align}

In conclusion, the criterion of symmetry invariance and unitary implementability of the quantum dynamics fixes, up to a global phase, the way in which the dominant terms in the asymptotic expansion of both $\tilde f_{\vec{k}}$ and $\tilde g_{\vec{k}}$ can depend on the unit wave vector $\vec{\tilde{k}}$ of the mode and on the time parameter (through the function $b_{\tilde k}$, which depends in turn on the vector $\vec{\tilde{k}}$ and on the Bianchi I directional scale factors).

\end{document}